\documentstyle[11pt,aaspp4]{article}

\slugcomment{Accepted by Astronomical Journal (December 2001 Issue)}

\begin{document}
\title{CCD PHOTOMETRY OF THE CLASSIC SECOND PARAMETER GLOBULAR CLUSTERS M3 AND M13
       \footnote{Data were obtained using the 2.4 m Hiltner Telescope of the
                  Michigan-Dartmouth-M.I.T. (MDM) Observatory.}\\}

\author{Soo-Chang Rey\altaffilmark{2}, Suk-Jin Yoon, Young-Wook Lee}
\affil{Center for Space Astrophysics \& Department of Astronomy,\\
    Yonsei University, Shinchon 134, Seoul 120-749, Korea\\
        Electronic mail : (screy, sjyoon, ywlee)@csa.yonsei.ac.kr}

\author{Brian Chaboyer}
\affil{Department of Physics and Astronomy, 6127 Wilser Lab, Dartmouth College, Hanover, NH 03755-3528\\
        Electronic mail : chaboyer@heather.dartmouth.edu}

\and

\author{Ata Sarajedini}
\affil{Department of Astronomy, 211 Bryant Space Science Center, P.O. Box 112055,\\
       University of Florida, Gainesville, FL 32611-2055\\
        Electronic mail : ata@astro.ufl.edu}

\altaffiltext{2}{Visiting Astronomer, MDM Observatory.}

\begin{abstract}
We present high-precision $V$, $B-V$ color-magnitude diagrams (CMDs) for the classic second
parameter globular clusters M3 and M13 from wide-field deep CCD photometry.
The data for the two clusters were obtained during the same photometric nights with the same 
instrument, allowing us to determine accurate relative ages. Based on a differential
comparison of the CMDs using the $\Delta (B-V)$ method, an age difference of 1.7 $\pm$ 0.7 Gyr
is obtained between these two clusters. We compare this result with our updated 
horizontal-branch (HB) population models, which confirm that the observed age difference
can produce the difference in HB morphology between the clusters. This provides further
evidence that age is the dominant second parameter that influences HB morphology.

\end{abstract}

\keywords{color-magnitude diagrams --- globular clusters: individual (M3, M13) 
           --- stars: evolution --- stars: horizontal-branch}

\section{INTRODUCTION}
Since the pioneering work by Sandage \& Wallerstein (1960), van den Bergh (1967), and 
Sandage \& Wildey (1967), it has been known that, in addition to [Fe/H] (the ``first parameter"),
there must be a second parameter controlling the morphology of the horizontal-branch (HB) in 
Galactic globular clusters (GGCs). Because of its important implications for the formation
chronology of the Galaxy (Searle \& Zinn 1978; Lee, Demarque, \& Zinn 1994, hereafter LDZ),
determining the nature of this second parameter has been one of the key questions during
the last forty years. The high-precision CCD photometry of GCs in recent years (Bolte 1989;
Buonanno et al. 1990, 1994; Green \& Norris 1990; Chaboyer, Demarque, \& Sarajedini 1996;
Sarajedini, Chaboyer, \& Demarque 1997; Sarajedini 1997; Stetson et al. 1999; Lee \& Carney 1999)
and the recent advances in HB modeling (LDZ) have suggested that age is most likely the cause of
the observed  variations in HB morphology among clusters of the same [Fe/H] and as a function of
Galactocentric distance. In particular, LDZ have concluded that age is the most natural candidate
for the global second parameter, because other candidates, such as helium abundance, CNO abundance,
and core rotation, can be ruled out from the observational evidence, while supporting evidence does
exist for the age hypothesis.

While the above view is generally accepted, some critics have argued that the relative age
differences inferred from the main-sequence turnoffs (MSTOs) are sometimes too small to explain
the differences in HB morphology between second parameter clusters (VandenBerg, Bolte, \& 
Stetson 1990, hereafter VBS; Stetson, VandenBerg, \& Bolte 1996; Catelan \& de Freitas Pacheco 
1995; Johnson \& Bolte 1998, hereafter JB98). Among them, the most problematic case is M3 and M13, 
which is one of the most famous second parameter pairs along with NGC 288 and NGC 362, where the
age difference is well established.
These studies argue that the age difference between M3 and M13 is appreciably smaller 
than the one suggested by HB population models, and therefore is not sufficient to explain
the observed difference in HB morphologies. The difference in HB morphology between M3 and M13,
however, is not as dramatic as that of NGC 362 and NGC 288, because M3, the cluster with the
redder HB, possesses an intermediate HB type with both blue and red HB stars. 
Therefore, the predicted age difference is smaller in the M3/M13 pair (see Lee et al. 1999 and below)
as compared with NGC 288/NGC 362, and is sometimes compatible with the observational uncertainties
(1 - 2 Gyr) of the available CCD photometry. 

Consequently, in order to test the age hypothesis of the second parameter effect between M3 and M13,
high-quality CCD data should be taken that is good enough to discriminate a small age difference.
For relative age dating of GGCs, it is essential to obtain CMDs that are reliable, at least,
in a differential sense. The inhomogeneity of datasets and analysis methods in the various
studies has been a major limitation of relative age dating of GCs. Until recently, many studies
combined CMDs obtained from different instruments with different calibrations. These results have
often been significantly hampered by this inhomogeneity and thus cannot be considered to be
conclusive (see review in Stetson et al. 1996; Rosenberg et al. 1999).
Indeed, even for the same GC, the fiducials that have been derived in different investigations
sometimes show large differences between them (e.g., see Fig. 12 of VandenBerg 2000 for M3 fiducials).
Consequently, for reliable relative age dating, the CCD frames for the two clusters must be taken 
during the same observing run using the same instruments, and also the same reduction and calibration
procedures must be employed for these frames.

Although M3 and M13, which have similar [Fe/H] values but different HB morphologies, provide a typical
example of the second parameter pair, there are only a few studies based on homogeneous datasets.
One of the best available CCD CMDs for M3 and M13 are provided by JB98. Their $VI$ photometry was 
obtained with the same telescope during the same observing run, and the resulting CMDs for the two 
clusters narrowly define the cluster sequences from the red giant branch (RGB) to $\sim$ 3 mag below
the MSTO. Using the differential age dating described by VBS, they concluded that the clusters are
unlikely to differ in age by the amount required to explain the differnece in their HB morphology
purely as an age effect. They proposed that the observations could be explained better with a
difference in the main-sequence (MS) helium abundance (with M13 having the larger helium abundance),
while this contradicts the observational evidence of similar helium abundances between M3 and M13
to within the errors ( Y = $0.204^{+0.011}_{-0.012}$ vs. $0.180^{+0.023}_{-0.027}$; Sandquist 2000).
On the other hand, Stetson (1998) obtained homogeneous $BI$ photometry for M3 and M13 using the CFHT,
and proposed that M13 appears to be about 12\% older than M3, although he argued that age is not
the only appropriate difference between M3 and M13, due to the bluer color of MSTO of M13 than
that of M3 and different slopes of subgiant branches (SGBs) in these two clusters.
From his preliminary instrumental $B, B-R$ CMD for M3 and M13, Sarajedini (1999) also presented a
result that M13 appears to be older than M3 by $\sim$ 2 Gyr and this age difference is consistent 
with that implied by HB models. Consequently, although there are at least three independent studies 
on the second parameter problem for M3 and M13 based on  homogeneous datasets, the situation is
still rather controversal.

The purpose of this paper is to present our new high-quality homogeneous $BV$ CCD photometry of 
M3 and M13. Our new photometry was obtained on the same photometric nights with the same instrument
producing a deeper and more extensive dataset as compared with that of JB98. This allows us to carry 
out a relative age dating for these clusters that is more precise than previous studies. These data are
then compared with our new updated HB models to test the age hypothesis of the second parameter effect
between M3 and M13.

\section{OBSERVATIONS AND DATA REDUCTIONS}
All the observations were made using the Michigan-Dartmouth-MIT (MDM) Observatory 2.4m 
telescope during two nights of an observing run in 2000 March. Images were obtained with
the thinned, back-illuminated, SITE 2048 ``Echelle" CCD and the standard Johnson $B$ and $V$
filters. The field of view of this CCD is roughly 9$\arcmin$.5 with a pixel scale of 0$\arcsec$.28.
We observed two partially overlapping fields in M3 and M13, respectively. One field was
about 12$\arcmin$ south-west (SW) of the cluster center in order to produce a deep and
well-defined MS in each cluster. Six long (200 - 800 s) and six short (20 - 60 s) exposures
were obtained at this location. In order to ensure a large sample of SGB and RGB stars,
a second field off the cluster center by 9$\arcmin$ west (W) was observed with three medium
(80 - 240 s) and three short (20 - 60 s) exposures. The nights were dark and of photometric quality,
and the seeing was also good - in the range 1$\arcsec$.0 - 1$\arcsec$.2. Figure 1 and Figure 2 show
the observed cluster fields for M3 and M13, respectively.

The raw data frames were calibrated using twilight and dawn sky flats and zero-level
exposures. Calibration frames were made by combining several individual exposures.
All exposure times were sufficiently long that the center-to-corner shutter timing
error was negligible. These procedures produced object frames with the sky flat
to better than 1\% in all filters.

Photometry of the M3 and M13 stars was accomplished using DAOPHOT II and  ALLSTAR
(Stetson 1987, 1995). For each frame, a Moffat function PSF, varying quadratically
with radial position, was constructed using 50 to 100 bright, isolated, and unsaturated stars.
The PSF was improved iteratively by subtracting faint nearby companions of the PSF stars.
We calculated aperture corrections using the program DAOGROW (Stetson 1990).
Using the aperture photometry data, growth curves were constructed for each frame,
in order to extrapolate from the flux measurements over a circular area of finite radius 
to the total flux observable for the star. The final aperture correction was made by
adjusting the ALLSTAR magnitude of each star by the weighted mean of the difference between
the total aperture magnitude returned by DAOGROW and the profile-fitting ALLSTAR magnitude 
for selected stars (e.g., PSF stars). The typical rms deviation for the aperture correction
for all frames corresponds to 0.013 mag, which introduces a modest uncertainty to the zero 
point of the calibration equation. After the aperture correction, DAOMATCH/DAOMASTER (Stetson 1992)
were then used to match stars of all frames covering the same field, and derive the average
instrumental magnitude on the same photometric scale. For each frame, the magnitude offset 
with respect to each master frame in $B$ and $V$ was calculated, and photometry for all
frames of the same field was transformed to a common instrumental system. In this way,
robust, intensity-weighted mean instrumental magnitudes and rms scatters of magnitudes 
were obtained for all matched stars. Last, the mean instrumental $B$ and $V$ magnitudes 
were matched to form $B-V$ color.

On each night, a number of standard stars from the list of Landolt (1992) were observed.
63 standard stars were observed in $B$ and $V$, covering a color range of --0.3 - 2.2
for $B-V$, and an airmass range of 1.16 - 1.46. All standard-star exposure times were long
($>$ 7 sec) enough so that the systematic error resulting from shutter timing ($\sim$ a few
tens of milliseconds) is insignificant ($<$ 1\%). DAOGROW was also performed to measure
the total aperture magnitude of the standard stars. The aperture magnitudes and the known
standard system magnitudes were then used to derive coefficients of the transformation
equations. The atmospheric extinction coefficients in each color have been determined by
the same standard stars at different airmasses. The final transformation equations were
obtained by a linear least-square fit. They are
        $$B-V = 1.130(b-v)_{o} + 0.280,$$
        $$V-v_{o} = -0.037(B-V) - 1.118,$$
where $B-V$ and $V$ are the color indices and visual magnitude in the standard $BV$ 
system, $(b-v)_{o}$ and $v_{o}$ refer to instrumental ones corrected for extinction.
No other trends in the residuals were noticeable, and therefore no additional terms
in the transformation equations appear to be necessary. The calibration equations
relate observed to standard values for $V$ and $B-V$ with rms residuals of
0.006 and 0.005 mag, respectively, as shown in Figure 3.

From comparison of the stars belonging to overlapping area of two adjacent fields,
we confirmed that there are no systematic differences in either color or magnitude between
the cluster fields. The mean offsets of the photometric zeropoint are never greater
than 0.01 mag.
We compared our CCD photometry with data taken from other studies for the stars in common.
Figure 4a and Figure 5a show comparison of our $V$ magnitudes with those of JB98 for M3 and M13
stars, respectively. The mean difference in the sense our measurements minus others is
0.002 $\pm$ 0.028 and 0.001 $\pm$ 0.024 for M3 and M13, respectively, where the uncertainty
is the standard deviation of the mean. Fig. 4b shows comparisons with photometry of
Stetson \& Harris (1988) for M3 (their secondary standard field). The mean differences are
0.026 $\pm$ 0.019 and 0.010 $\pm$ 0.022 in $V$ and $B-V$, respectively. We have also made a 
comparison with the photometry of Richer \& Fahlman (1986) for M13, as shown in Fig. 5b.
The mean differences are 0.019 $\pm$ 0.019 and 0.009 $\pm$ 0.014 in $V$ and $B-V$, respectively.
We conclude that there are no significant systematic zero-point differences between our photometry
and others.

\section{COLOR-MAGNITUDE DIAGRAMS}
For the final photometric list, we selected stars with a detection on at least three 
frames in each band, error in the $B$ and $V$ magnitudes of less than 0.05 mag, and
a mean value of $\chi$ (quality of the PSF fit to the stellar image returned by DAOPHOT II)
less than 2. Figure 6 shows our $V$, $B-V$ CMDs for the M3 and M13, respectively, 
representing stars used in our analysis. All of the long exposure data for the SW field
of M3 and M13 were included except saturated bright RGB stars which were recovered from
the short exposure data. However, in the case of the W field data, the following selection
criteria were adopted to have the best definitions of the CMD branches since crowding
worsens the quality of the photometry in the inner cluster region. We included only stars
with $V <$ 18 mag from the medium exposure data of M3. For the sparsely populated bright
RGB and SGB of M13, we added medium and short exposure photometry for the relatively bright 
stars ($V <$ 18.7, $B-V >$ 0.3) in outer part of the CCD frame (Y $>$ 700 pixel) to help
define the sequences more accurately. 
In order to represent the HB blue tail of M13 which extends to about the MSTO magnitude,
we include all medium exposure data with $B-V <$ 0.3. Note that the CMD of M13 is being
used here to understand the morphology of the CMD and does not accurately represent 
the population ratios of stars in different evolutionary stages. The CM data for M3 and M13
are tabulated in Table 1 and Table 2, respectively. In each table, most data are taken
from the SW field, while 5000 series identifications are from the W field.

Both our M3 and M13 CMDs extend to about 3.5 magnitudes fainter than the MSTO.
All of cluster sequences that we are particularly interested in (the lower RGB, SGB, MS, 
and MSTO) are significantly well defined, allowing us to derive the cluster parameters
accurately. Our CMDs show the similar quality and overall morphology with those of previous
investigators (e.g., JB98), but include more stars in all the cluster sequences. Although there 
is a paucity of stars in the brightest RGB and asymptotic giant branch regions in both cluster
CMDs, this does not hamper our differential age analysis which mainly uses the lower RGB stars.

M13 has an exclusively blue HB with a long blue tail extending well below the level of the MSTO.
Moreover, the HB appears to have two gaps at $V \sim$ 15.6 and $V \sim$ 17.6. These gaps
are also presented in the previous $BV$ data (Paltrinieri et al. 1998) and $HST$ UV/visual data
(Ferraro et al. 1997, 1998) at similar locations to ours. Ferraro et al. (1998) suggested
that the gaps are not statistical fluctuations, but real, because gaps seem to be at similar 
locations (i.e., at similar $T_{eff}$) in different clusters (see also Piotto et al. 1999;
D'Cruz et al. 2000). The two apparent gaps in our HB of M13 appear to be the G1 and G3 gaps 
labeled by Ferraro et al. (1998).

The fiducial line of the MS and RGB was constructed by iteratively rejecting a few stars 
which deviate by more than 3$\sigma$ from the main branches, binning stars over 0.2 mag 
intervals in $V$, and then determining the robust mean value in $V$ and $B-V$ for each bin.
The MSTO region, which has a strong curvature and is critical for differential age dating 
techniques, was sampled with a smaller magnitude bin of 0.1 mag. Because of the very small
number of potential members brighter than $V\sim$14.5 and $V\sim$13.5 for M3 and M13, respectively,
the fiducial line of bright RGB sequence is determined by eye. Our unsmoothed M3 and M13
fiducial sequences are listed in Table 3 and Table 4, respectively. 
We found no significant difference between RGB fiducials from the W and SW fields.
There is some disagreement of our fiducial for M13 with that of Paltrinieri et al. (1998). 
However, we found that our fiducial has good agreement with those obtained from other studies
such as Richer \& Fahlman (1988) and Yim et al. (2000). While these two fiducials agree well
with ours in the region of MS and SGB, Paltrinieri et al.'s fiducial shows large deviation
in this region.

\section{RELATIVE AGE DATING: $\Delta(B-V)$ METHOD}

In contrast to absolute age dating techniques, relative age determinations are less affected
by uncertainties in stellar evolution theory, because most of the effects of theoretical
uncertainties are removed in a relative comparison. Furthermore, differential comparisons
between GCs can reduce the effects of observational uncertainties involved in the absolute
age determination (see Stetson et al. 1996; Sarajedini et al. 1997). For example, VBS and
Sarajedini \& Demarque (1990) independently showed that, for clusters having similar
chemical compositions, the color difference between the MSTO and the lower giant branch
(and the SGB) is an excellent age diagnostic, in the sense that a larger color difference
denotes a younger age; one major advantage of this diagnostic is that it is independent of 
distance, reddening, and photometric zero-point/color calibrations. The technique outlined by 
VBS involves normalizing cluster principal sequences by the TO color, $(B - V)_{TO}$,
and the visual magnitude, $V_{+0.05}$, of the MS at the point 0.05 mag redder than the TO color.
Once this shifting has been accomplished, older clusters will have bluer giant branches
relative to younger clusters and vice versa.

The $(B - V)_{TO}$ and $V_{+0.05}$ were determined by fitting a parabola to the stars in the
region near the MSTO and upper MS point 0.05 mag redder than the TO, respectively.
We derived $(B - V)_{TO}$ = 0.440 $\pm$ 0.005 and 0.435 $\pm$ 0.006 for M3 and M13, respectively, 
where the uncertainty is the observed scatter of the stars in the TO region. The $V_{+0.05}$ was
also obtained to be 19.91 $\pm$ 0.06 and 19.49 $\pm$ 0.06 for M3 and M13, respectively, where the
error is also the observed scatter of the stars within the parabola-sided boxes including $V_{+0.05}$.

In Figure 7, we show our CM data and fiducial sequences for M3 and M13 using the registration
prescription specified by VBS. In Fig. 7b, also shown are Yonsei-Yale (Y${^2}$) isochrones
(Yi et al. 2001) having [Fe/H] = -1.66 
\footnote{
Our models reveal that the $\Delta$ Age/$\Delta (B-V)$  of RGBs will not be affected
by choice of [Fe/H] from -1.9 to -1.4.}
(Zinn \& West 1984) and [$\alpha$/Fe] = 0.3 and ages ranging 
from 10 to 14 Gyr. The M3 and M13 sequences and the isochrones have been shifted horizontally 
to match at $(B - V)_{TO}$, and vertically to agree with one another at $V_{+0.05}$.
As shown in the figures, the RGB sequences of M3 and M13 are well separated from each other,
indicating a real age difference.

It is important to note that, in Fig. 7, the separation in the SGB region between M3 and M13 appears
to be inconsistent with the original registration scheme of VBS, where the SGBs of the two clusters 
should be coincident. As shown in the isochrones of Fig. 7b, this separation is partly explained as 
a real feature if the absolute ages of the two clusters are as young as 10 - 12 Gyr. This feature was 
also found in the $VI$ CMDs reported by JB98. However, their CMDs show a larger separation between
the SGBs of the two clusters than ours, relative to the separation of the RGBs caused by an age
difference (see Fig. 8 and Fig. 9 of JB98). VandenBerg (2000) found that this is due to the failure
of JB98 to match their data in the TO region, and explained that matching the CMDs at $V_{+0.05}$
will not produce a superposition of the cluster TOs if the MS loci have different slopes.

In Fig. 7, we also see that the TO of M13 is slightly brighter than that of M3 after the cluster
fiducials have been registerd following the VBS method. Following the recommended
procedures of VandenBerg (2000, see his footnote 3), we therefore shifted our M13 fiducial
vertically until the differences between the two clusters are minimized at the two points that
are 0.05 mag redder than the TO (i.e., both above and below the TO). The M13 sequence has been
shifted fainter and bluer by 0.06 mag and 0.001 mag, respectively. In Figure 8, we present 
our M3 and M13 fiducial sequences and individual CM data, after the TOs of both clusters are
made to coincide with each other, along with the same isochrones shown in Fig. 7b. It should be noted
that there is still a sufficiently large separation between the M3 and M13 RGBs, implying an
age difference.

The relative age is measured by comparing the colors of the RGB, where the color separation is
largely independent of magnitude. In the range of -4 $< V - V_{+0.05} <$ -2, we fit a parabola
to the M3 RGB sequence and calculated the mean offset between this parabola and the shifted RGB 
stars of M13. In order to check the reliability of this estimate, we repeated this procedures for 
the case of the M13 RGB sequence and the M3 stars. We find the mean value of the color difference
between the RGBs, in the sense of M3 - M13, to be $\delta$$(B - V)_{RGB}$ = 0.029 $\pm$ 0.008.
The error is the standard error of the mean color difference. 
We estimated $\sigma$ $(B-V)$ = 0.008 in the separation of the RGBs due to the error of the MSTO color.
Considering the additional error (0.08 mag) from the magnitude of the MSTO which 
corresponds to the $\sigma$ $(B-V)$ = 0.004 in the RGBs, we estimated $\sigma$ $(B-V)$ = 0.009
due to the combined error in the magnitude and color of MSTO. When this error is combined with 
that of the mean color difference between the RGBs, the total uncertainty of the color difference
between the RGBs of M3 and M13 corresponds to 0.012 mag.

We calculate the offset in RGB
color associated with an offset in presumed age by noting that $\Delta(log t_{9})$ = -2.19$\Delta(B-V)$. 
This corresponds to about 0.59 Gyr per 0.01 mag, if we adopt an age of 12 Gyr for M13.
It should be noted that the color difference between the RGBs is not uniform, but depends on the
absolute age (see the spacing of isochrones presented in Fig. 7 and 8), in the sense that using 
younger isochrones would reduce the inferred age differences. Adopting the absolute age of M13 to
be about 12 Gyr (Chaboyer et al. 1998; Yi et al. 2001), the color difference between the two clusters
corresponds to a relative age difference of 1.7 $\pm$ 0.7 Gyr, in the sense that M13 is older.
\footnote{
Note that a difference in [O/Fe] is unlikely to account for the difference in $\Delta(B-V)$
between M3 and M13. This is because M13 would need to have a higher [O/Fe] value ($\sim$ 0.8 dex 
at [Fe/H] = -1.6 and 12 Gyr, see VandenBerg \& Stetson 1991) than M3 and if this is the case,
then the HB of M13 would be redder than that of M3 (LDZ). This situation is the opposite of
what is observed in these two clusters. On the other hand, from a measurement of the oxygen
abundance in relatively less evolved stars near the base of the RGB (i.e., in the vicinity
of the first dredgeup), Pilachowski \& Armandroff (1996) suggested that M13 is as much as
$\Delta$[O/Fe] $\sim$ 0.3 dex more oxygen poor than M3, while this is still a disputable result
(see Kraft et al. 1997). If true, this oxygen abundance difference will increase the inferred 
age difference ($\sim$ 0.7 Gyr, see VandenBerg \& Stetson 1991), since the effects of decreasing
[O/Fe] are to increase the $\Delta(B-V)$ of M13 at a given age.}$^,$
\footnote{
In order to check the dependence of derived age difference on the part of the RGB
which is considered, we derived age difference between M3 and M13 in the range of
-3 $< V-V_{0.05} <$ -2 and -4 $< V-V_{0.05} <$ -3, separately. We find the mean
value of the color difference between RGBs, in the sense of M3 - M13, to be
$\delta$$(B - V)_{RGB}$ = 0.027 $\pm$ 0.008 and 0.030 $\pm$ 0.008, respectively.
From the isochrones, we also calculate $\Delta(log t_{9})$ = -2.39$\Delta(B-V)$ 
and $\Delta(log t_{9})$ = -2.08$\Delta(B-V)$, respectively. This corresponds to
about 0.64 Gyr and 0.56 Gyr per 0.01 mag. Therefore, the color differences between
the two clusters correspond to the relative age differences of 1.73 Gyr and 
1.68 Gyr, respectively, which show good agreement with our adopted 
relative age difference (1.7 Gyr).}

\section{COMPARISON WITH SYNTHETIC HORIZONTAL-BRANCH MODELS}
There are several recent developments that affect the relative age dating from HB morphology.
We have included them in our most updated version of the HB population models (see Lee et al. 1999;
Lee \& Yoon 2001, in preparation). First of all, there is now reason to believe that the absolute ages
of the oldest GGCs is reduced to about 12 Gyr, as suggested by the Hipparcos distance calibration
and other improvements in stellar models (Reid 1998; Gratton et al. 1997; Chaboyer et al. 1998;
Grundahl, VandenBerg, \& Andersen 1998; Yi et al. 2001). This has a strong impact on the relative 
age estimation from HB morphology,
because the variation of the HB mass is more sensitive to age at younger ages, due to the
nonlinear relationship between RGB tip mass ($M_{RG}$) and age (see LDZ). Second, in the new 
population models, we have used new HB tracks with improved input physics (Yi et al. 1997) 
and corresponding Y${^2}$ isochrones (Yi et al. 2001). Third, it is now well established that the
$\alpha$-elements are enhanced in halo populations. Specifically, we adopted [$\alpha$/Fe] = 0.3
for clusters with [Fe/H] $<$ -1.0, and thereafter we assume that it steadily declines to 0.0
at solar metallicity (e.g., Wheeler, Sneden, \& Truran 1989). In practice, the treatment suggested
by Salaris, Chieffi, \& Straniero (1993) was used to simulate the effect of $\alpha$-element
enhancement. Finally, empirical mass-loss law of Reimers (1975) suggests more mass-loss at
larger ages. The result of this effect was also presented in LDZ, but unfortunately their
most widely used diagram (their Fig. 7) is the one based on fixed mass-loss.

We found that all of the above effects make the HB morphology more sensitive to age.
Figure 9 illustrates the HB morphology vs. [Fe/H] relations for the GGCs, along with the theoretical
HB isochrones that were produced by our synthetic HB models. As shown in Fig. 9b, now the required
age difference is significantly reduced as compared to LDZ. Only 1.1 Gyr of age difference, 
rather than 2 Gyr, is enough to explain the systematic shift of the HB morphology between the inner 
and outer halo clusters.
To within the observational uncertainties, age differences of about 1 - 2 Gyr are now enough to
explain the observed differences in HB morphology between M3 and M13 (or M2), and between the outer
halo clusters (Pal 3, Pal 4, Pal 14, and Eridanus) and M3. These values are consistent with the 
observations presented in this paper and also with the recent relative age datings both from 
HST and high-quality ground-based data (Stetson et al. 1999; Lee \& Carney 1999).

In Figure 10, the observed CMDs of M3 and M13 are compared with our new population models,
which include the scatter expected from the random errors in magnitude and in color as
estimated by our photometry. For the two clusters, we adopted the same metallicity ([Fe/H] = -1.66)
on the Zinn \& West (1984) scale,
mass dispersion ($\sigma_{M}$ = 0.025 M$_{\odot}$) on the HB, MS helium abundance ($Y_{MS}$ = 0.23), 
and $\alpha$-element enhancement ([$\alpha$/Fe] = 0.3), but applied an age difference of 1.7 Gyr 
between the two clusters as estimated from our relative age dating (see Sec. 4). As shown in the
figure, there is a reasonable match between the synthetic HB models and the observations. 
However, the observations indicate that M13 has a long blue tail on the HB, while the standard HB model
(with $\sigma_{M}$ $\sim$ 0.02 - 0.03 M$_{\odot}$) fails to reproduce this detail.

The wide color range of the M13 HB would be reproduced by using a larger value for $\sigma_{M}$, roughly
twice as large as M3, together with a slightly larger age difference ($\Delta$t $\sim$ 2.4 Gyr) between
the two clusters (see Lee \& Yoon 2001, in preparation). Certainly, this uncertainty has some impact on
the age difference that one infers from HB morphology. However, the magnitude of this age uncertainty
($\sim$ 0.7 Gyr) is still compatible with the errors ($\sim$ 0.7 Gyr) of the relative age dating
from MSTO photometry and/or additional age difference ($\sim$ 0.7 Gyr) due to the possible difference of 
[O/Fe] between two clusters (see footnote 4). This suggests that the difference in age between M3 and M13 
inferred from their MSTOs can account for most of the difference in the HB distribution between the
two clusters. Hence, age appears to be predominantly responsible for the second parameter effect 
in the M3/M13 pair despite the current uncertainty about the origin of M13's blue HB extension
(see Sec. 6 for discussion).

\section{DISCUSSION}

Although the presence of a blue tail on the HB has only a mild impact on the relative age dating 
from HB morphology, it is still important to understand the origin of this effect in order to use
the HB as a more reliable age indicator. The blue tail phenomenon is widely considered to be a
result of local effects, such as enhanced mass-loss in high density environments.
Buonanno et al. (1997) examined the role of stellar density in the morphology of the HB, and
suggested that clusters with higher central densities are more likely to populate the bluest
extremes of the HB (see also Fusi Pecci et al. 1993). However, the correlation between stellar
density and HB morphology is rather weak with large scatter and/or limited to a small
fraction of GCs within an intermediate metallicity range (LDZ; Sarajedini et al. 1997).
Note that the central densities (log $\rho$ = 3.51 vs. 3.33, Harris 1996) and concentration 
(c = log $r_{t}$/$r_{c}$ = 1.84 vs. 1.51, Harris 1996) parameters for M3 and M13 are very 
similar despite their apparent difference in $\sigma_{M}$ on the HB. Furthermore, as suggested
by LDZ, there is also no clear evidence that the variation of HB morphology with galactocentric
distance is related to central densities of GCs.

It is also suggested that some noncanonical effects in the stellar interior, such as rapid rotation
and deep mixing would make a star both bluer and brighter on the HB, and are thus related to the
presence of blue tails (Mengel \& Gross 1976; Kraft et al. 1993, 1997; Kraft 1994; 
Peterson, Rood, \& Crocker 1995; Langer \& Hoffmann 1995; Sweigart 1997a, 1997b; Cavallo \& Nagar 2000).
However, the predicted increase in the rotation velocity with effective temperature along the
HB has not been confirmed from the recent high-resolution spectroscopy of blue HB stars of M13
(Behr et al. 2000). On the other hand, Behr et al. (1999) reported that, in the hotter HB stars 
of M13, helium is underabundant, while iron and other metals are enhanced. It is suggested that
these abundance anomalies are most likely due to the diffusion effects in the radiative atmospheres.
Similarly, a number of interesting phenomena have recently been reported to occur in blue HB 
stars around a temperature of 11,000 K; these include a gap (i.e., G1 gap) in the HB distribution
(Ferraro et al. 1998; Piotto et al. 1999; Caloi 1999), a jump in the Str$\ddot o$mgren $u$ magnitudes and 
the onset of radiative levitation (Grundahl et al. 1998, 1999), and a shift to lower surface gravities
(Moehler et al. 1999, 2000). All of these phenomena suggest the blue tail feature may be related to
the disappearance of surface convection and the formation of a radiative stellar atmosphere followed by
radiative levitation of heavy elements and helium diffusion for stars hotter than about 11,000 K
(Sweigart 2001; Moehler et al. 1999). On the theoretical point of view,
it is probably possible that the levitation of heavy elements along with
helium diffusion would push the blue HB stars to even hotter $T_{eff}$ on the HR diagram,
creating the blue tail. If this is confirmed by detailed modeling, the blue tail phenomenon
may not be considered as adding noise to the second parameter effect, since it is then rather
a general feature of extremely blue HB clusters. In this case, a more reasonable relative age
would be estimated from the HB morphology by ignoring the blue tail, since radiative levitation
and diffusion effects are not included in our standard HB models (i.e., Fig. 10).

\section{SUMMARY}

We present new high-quality $V$, $B-V$ CMDs for the Galactic globular clusters 
M3 and M13 constructed from wide-field deep CCD photometry obtained during the
same nights with the same instrument. From our homogeneous dataset, we draw the 
following conclusions:
\\1. Based on a careful differential comparison of the CMDs using the $\Delta (B-V)$
method, we confirm a significant difference between these two clusters,
indicating an age difference of 1.7 $\pm$ 0.7 Gyr in the sense that M13 is older than M3.
\\2. We present updated HB models, which suggest that HB morphology is more sensitive
to cluster age compared to our previous models. From a comparison of observations with
the new HB models, we find that the observed age difference can reproduce the difference
in HB morphology between the clusters. This provides further evidence that cluster age is the
dominant second parameter that influences HB morphology, which in turn suggests that HB
morphology is a reliable age indicator in most population II stellar systems.
\\3. While the physical origin of the blue tail phenomenon is still uncertain, there is 
now a growing body of evidence that suggest this is an ubiquitous characteristic of
clusters with extremely blue HB stars hotter than 11,000 K. If true, the presence of
blue tail would have less impact on  relative age dating based on HB morphology.

\acknowledgments
We would like to thank J. Johnson for providing electronic copy of data and an
anonymous referee for a careful review and useful comments.
Support for this work was provided by the Creative Research Initiative Program of
Korean Ministry of Science \& Technology. B. C.'s research is supported in part by
NASA grant NAG5-9225. A. S. acknowledges financial support from a National Science
Foundation CAREER grant No. AST-0094048.

\clearpage
\begin{deluxetable}{lrrcccc}

\setcounter {table}{0}
\footnotesize
\tablewidth{0pt}
\tablecaption{CCD Photometry for M3{*}}
\tablehead{
\colhead{ID}  & \colhead{$x$}  &   \colhead{$y$}  &   \colhead{$V$}   &   \colhead{$\sigma_{V}$}
&    \colhead{$B-V$}   &    \colhead{$\sigma_{(B-V)}$}}
\startdata
        1    & 1467.63    & 1106.58    &   11.406    &    0.005    &    0.702    &    0.007 \nl
        2    &  644.02    &   26.79    &   12.853    &    0.004    &    1.426    &    0.006 \nl
        3    &  241.15    &  354.39    &   14.047    &    0.003    &    0.664    &    0.005 \nl
        4    &  285.23    & 1191.12    &   14.165    &    0.003    &    1.024    &    0.005 \nl
        5    & 1559.35    & 1816.76    &   14.840    &    0.003    &    0.804    &    0.005 \nl
        6    &  400.20    & 1052.79    &   15.180    &    0.003    &    1.119    &    0.005 \nl
        7    &  286.05    &  484.68    &   15.349    &    0.003    &    0.851    &    0.005 \nl
        8    &  313.97    &  470.24    &   15.392    &    0.006    &    0.389    &    0.008 \nl
        9    &   99.29    &  227.80    &   15.415    &    0.005    &    0.550    &    0.007 \nl
       10    &  835.15    & 1745.64    &   15.540    &    0.006    &    0.820    &    0.008 \nl
       ..    &   .....    &   .....    &    .....    &    .....    &    .....    &    ..... \nl
       ..    &   .....    &   .....    &    .....    &    .....    &    .....    &    ..... \nl
       ..    &   .....    &   .....    &    .....    &    .....    &    .....    &    ..... \nl
     5001    &  640.73    &  649.51    &   12.741    &    0.023    &    1.415    &    0.024 \nl
     5002    & 1505.17    &  209.86    &   13.510    &    0.006    &    1.188    &    0.009 \nl
     5003    & 1847.97    &  271.11    &   14.019    &    0.008    &    1.057    &    0.010 \nl
     5004    & 1155.44    &   72.02    &   14.265    &    0.006    &    0.909    &    0.009 \nl
     5005    &  930.84    & 1552.24    &   14.538    &    0.006    &    0.978    &    0.008 \nl
       ..    &   .....    &   .....    &    .....    &    .....    &    .....    &    ..... \nl
       ..    &   .....    &   .....    &    .....    &    .....    &    .....    &    ..... \nl
       ..    &   .....    &   .....    &    .....    &    .....    &    .....    &    ..... \nl
\enddata
\tablenotetext{*} {Table 1 is presented in its complete form in the electronic edition of the
                  Astronomical Journal. The first page of this table is presented here for
                  guidance regarding its form and content.}
\end{deluxetable}

\clearpage
\begin{deluxetable}{lrrcccc}

\setcounter {table}{1}
\footnotesize
\tablewidth{0pt}
\tablecaption{CCD Photometry for M13{*}}
\tablehead{
\colhead{ID}  & \colhead{$x$}  &   \colhead{$y$}  &   \colhead{$V$}   &   \colhead{$\sigma_{V}$}
&    \colhead{$B-V$}   &    \colhead{$\sigma_{(B-V)}$}}
\startdata
        1    &  466.76    &  880.00    &   14.242    &    0.005    &    0.573    &    0.007 \nl
        2    &  322.68    & 1512.17    &   14.308    &    0.004    &    0.889    &    0.006 \nl
        3    &  144.05    &  115.67    &   14.503    &    0.004    &    0.844    &    0.006 \nl
        4    & 1373.81    &  235.60    &   14.721    &    0.004    &    0.597    &    0.006 \nl
        5    &   54.15    &  183.29    &   14.736    &    0.004    &    0.824    &    0.006 \nl
        6    &  453.66    &  688.43    &   14.889    &    0.004    &    0.581    &    0.006 \nl
        7    & 1697.89    &  520.95    &   15.030    &    0.004    &    1.379    &    0.006 \nl
        8    &  400.83    &   97.12    &   15.045    &    0.004    &    0.788    &    0.006 \nl
        9    &  893.09    &  604.51    &   15.063    &    0.004    &    0.073    &    0.006 \nl
       10    &   98.33    &  687.16    &   15.076    &    0.004    &    0.076    &    0.006 \nl
       ..    &   .....    &   .....    &    .....    &    .....    &    .....    &    ..... \nl
       ..    &   .....    &   .....    &    .....    &    .....    &    .....    &    ..... \nl
       ..    &   .....    &   .....    &    .....    &    .....    &    .....    &    ..... \nl
     5001    &  530.49    &  325.30    &   11.973    &    0.007    &    1.552    &    0.009 \nl
     5002    & 1571.01    &  281.42    &   12.203    &    0.006    &    1.306    &    0.009 \nl
     5003    & 1024.98    &  160.02    &   12.488    &    0.007    &    1.297    &    0.009 \nl
     5004    &  647.51    &   16.39    &   12.668    &    0.008    &    1.182    &    0.010 \nl
     5005    & 1518.46    &  258.04    &   12.708    &    0.008    &    1.200    &    0.010 \nl
       ..    &   .....    &   .....    &    .....    &    .....    &    .....    &    ..... \nl
       ..    &   .....    &   .....    &    .....    &    .....    &    .....    &    ..... \nl
       ..    &   .....    &   .....    &    .....    &    .....    &    .....    &    ..... \nl
\enddata
\tablenotetext{*} {Table 2 is presented in its complete form in the electronic edition of the
                  Astronomical Journal. The first page of this table is presented here for
                  guidance regarding its form and content.}
\end{deluxetable}

\clearpage
\begin{deluxetable}{cccc}

%\tablenum{1}
\setcounter {table}{2}
\footnotesize
\tablewidth{0pt}
\tablecaption{M3 fiducial sequence}
\tablehead{
\colhead{$V$}   &    \colhead{$B-V$}   &  \colhead{$V$}   &    \colhead{$B-V$}}
\startdata
14.72 & 0.928 & 19.50 & 0.456 \nl
15.24 & 0.863 & 19.71 & 0.474 \nl
15.83 & 0.806 & 19.89 & 0.489 \nl
16.31 & 0.759 & 20.10 & 0.507 \nl
16.80 & 0.727 & 20.30 & 0.538 \nl
17.09 & 0.712 & 20.51 & 0.562 \nl
17.25 & 0.699 & 20.70 & 0.595 \nl
17.53 & 0.688 & 20.91 & 0.627 \nl
17.69 & 0.669 & 21.10 & 0.668 \nl
17.89 & 0.669 & 21.30 & 0.709 \nl
18.11 & 0.644 & 21.50 & 0.749 \nl
18.31 & 0.614 & 21.70 & 0.798 \nl
18.49 & 0.520 & 21.91 & 0.846 \nl
18.69 & 0.457 & 22.09 & 0.873 \nl
18.91 & 0.445 & 22.29 & 0.895 \nl
19.10 & 0.438 & 22.48 & 0.941 \nl
19.30 & 0.447 & ..... & ..... \nl

\enddata
\end{deluxetable}

\clearpage
\begin{deluxetable}{cccc}

%\tablenum{2}
\setcounter {table}{3}
\footnotesize
\tablewidth{0pt}
\tablecaption{M13 fiducial sequence}
\tablehead{
\colhead{$V$}   &    \colhead{$B-V$}   &  \colhead{$V$}   &    \colhead{$B-V$}}
\startdata
13.75 & 0.943 & 18.68 & 0.436 \nl
14.25 & 0.876 & 18.90 & 0.444 \nl
14.76 & 0.821 & 19.10 & 0.457 \nl
15.25 & 0.775 & 19.31 & 0.469 \nl
15.76 & 0.734 & 19.51 & 0.485 \nl
16.24 & 0.701 & 19.69 & 0.508 \nl
16.77 & 0.668 & 19.90 & 0.531 \nl
17.13 & 0.654 & 20.10 & 0.564 \nl
17.29 & 0.639 & 20.31 & 0.603 \nl
17.49 & 0.631 & 20.50 & 0.637 \nl
17.70 & 0.613 & 20.69 & 0.678 \nl
17.90 & 0.532 & 20.90 & 0.712 \nl
18.11 & 0.464 & 21.10 & 0.763 \nl
18.29 & 0.443 & 21.30 & 0.800 \nl
18.51 & 0.436 & 21.49 & 0.837 \nl
\enddata
\end{deluxetable}

\clearpage

\begin{figure}
\plottwo{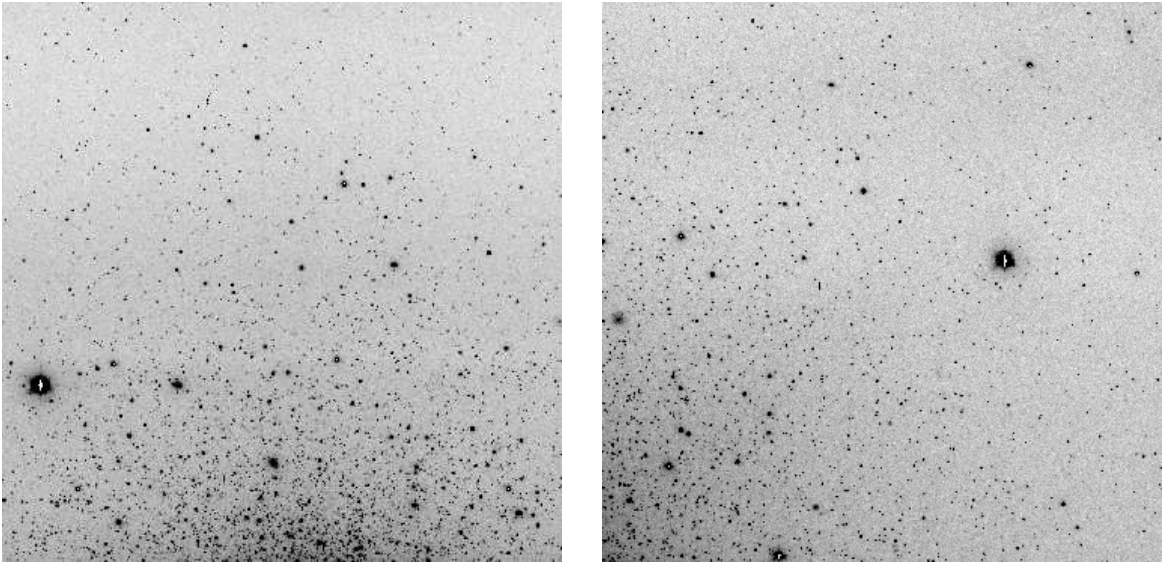}{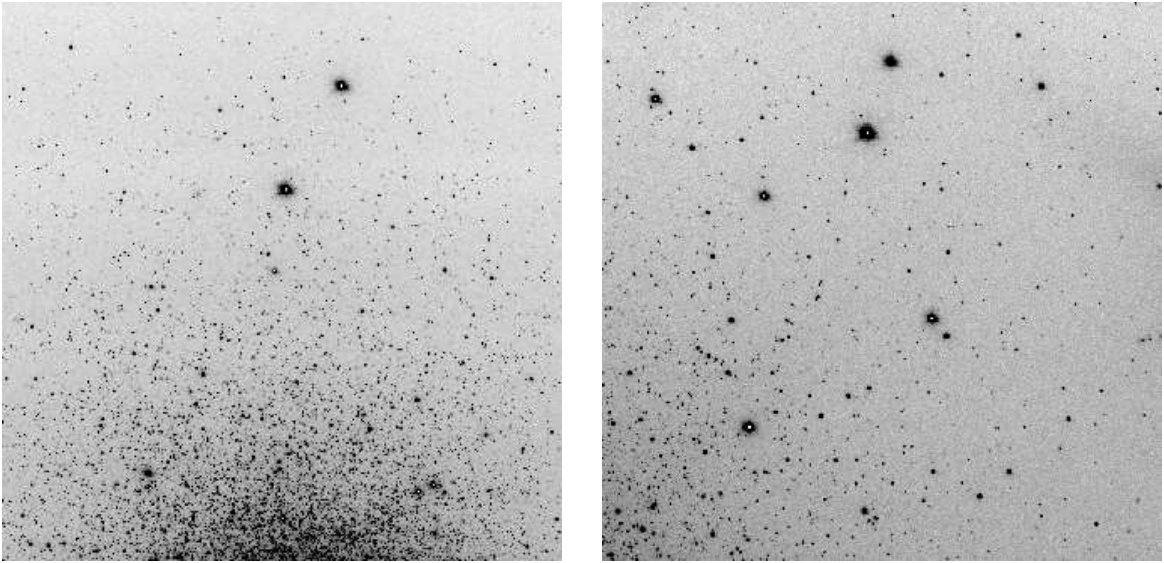}
 \caption{Observed cluster regions in (a) west and (b) south-west field of M3.
          North is to the left and east is down.}
\end{figure}

\begin{figure}
\plottwo{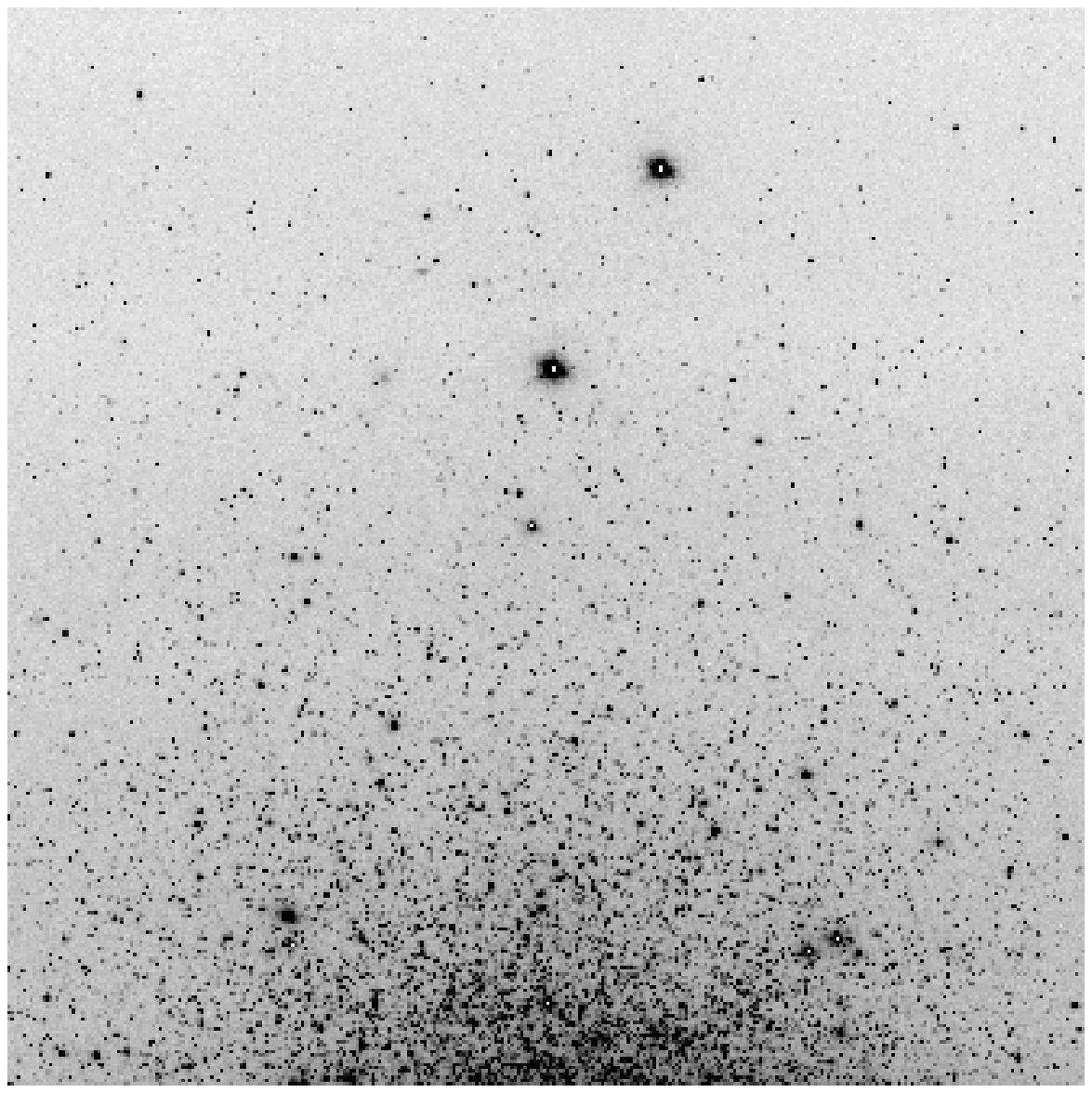}{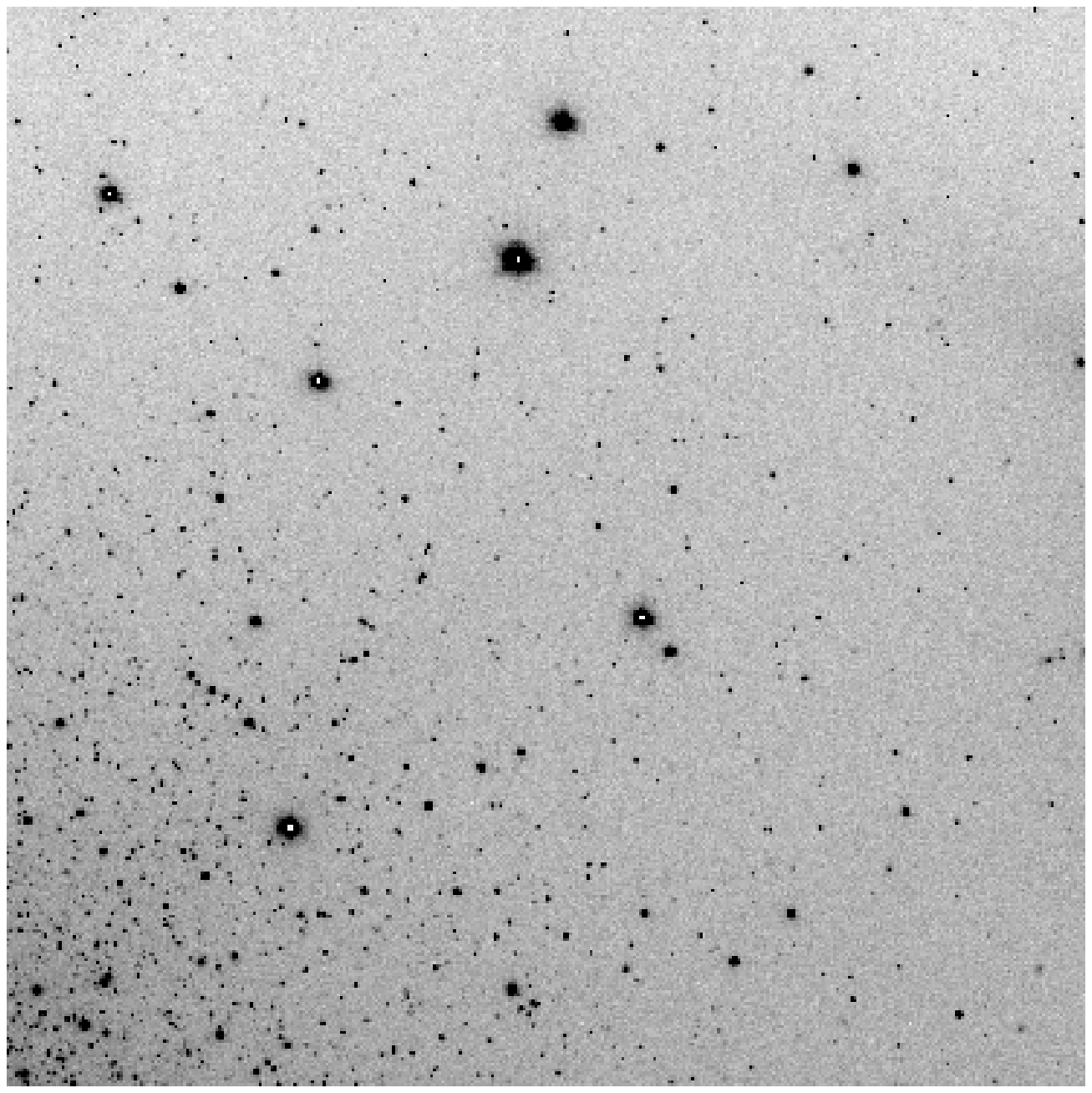}
 \caption{Observed cluster regions in (a) west and (b) south-west field of M13.
          North is to the left and east is down.}
\end{figure}

\begin{figure}
\centerline{\epsfysize=7in%
\epsffile{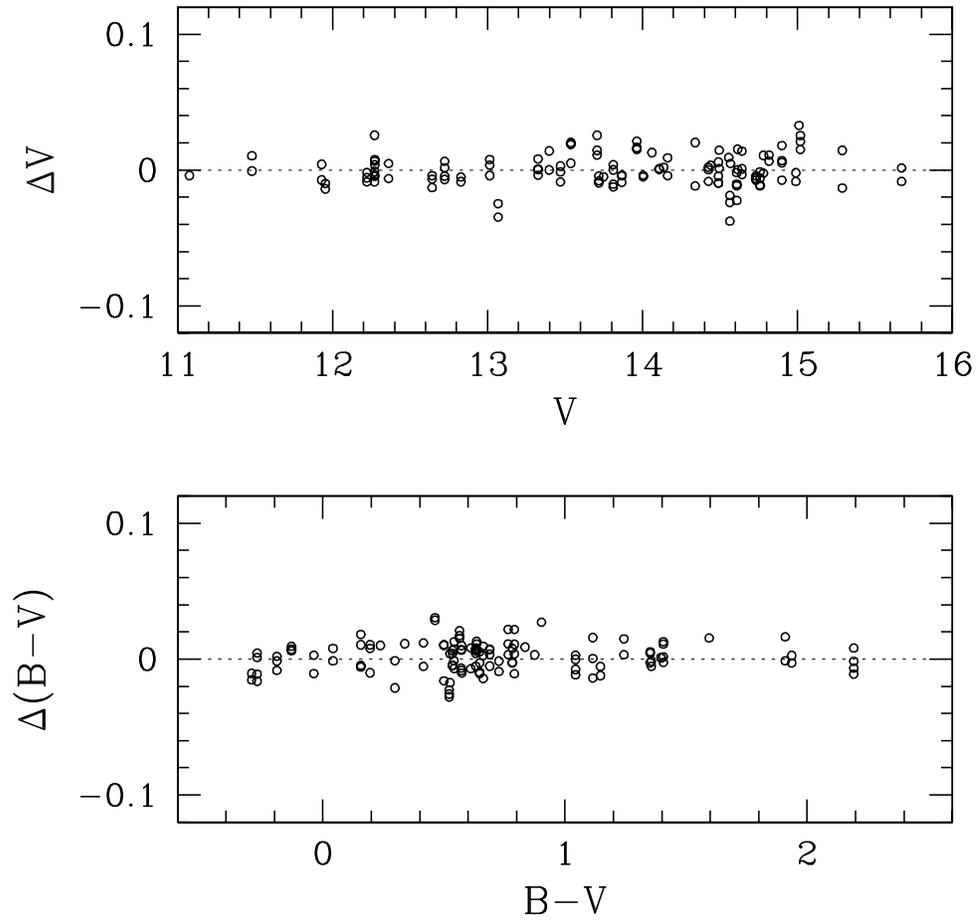}}
 \caption{Magnitude and color residuals for the comparison of Landolt (1992) standards and
measured values in this study, in the sense of us minus Landolt.}
\end{figure}

\begin{figure}
\centerline{\epsfysize=7in%
\epsffile{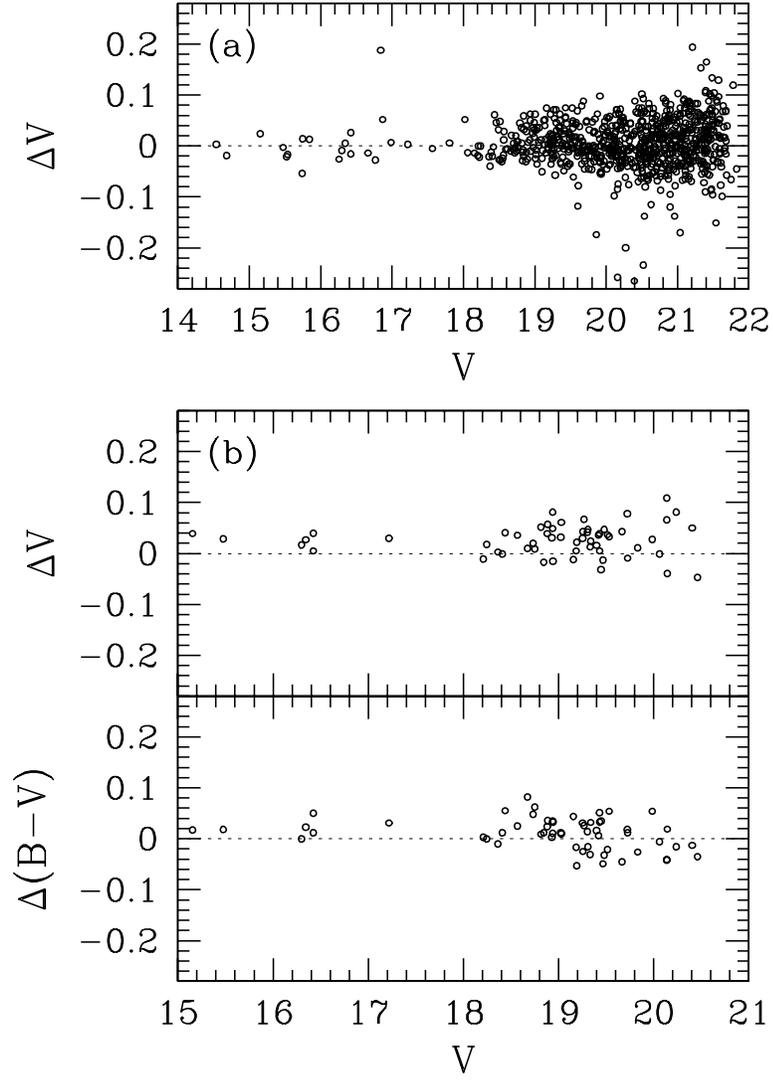}}
 \caption{Comparison of our photometry with (a) Johnson \& Bolte (1998) and (b) Stetson \& Harris (1988)
for M3. The differences are in the sense of our photometry minus others.}
\end{figure}

\begin{figure}
\centerline{\epsfysize=7in%
\epsffile{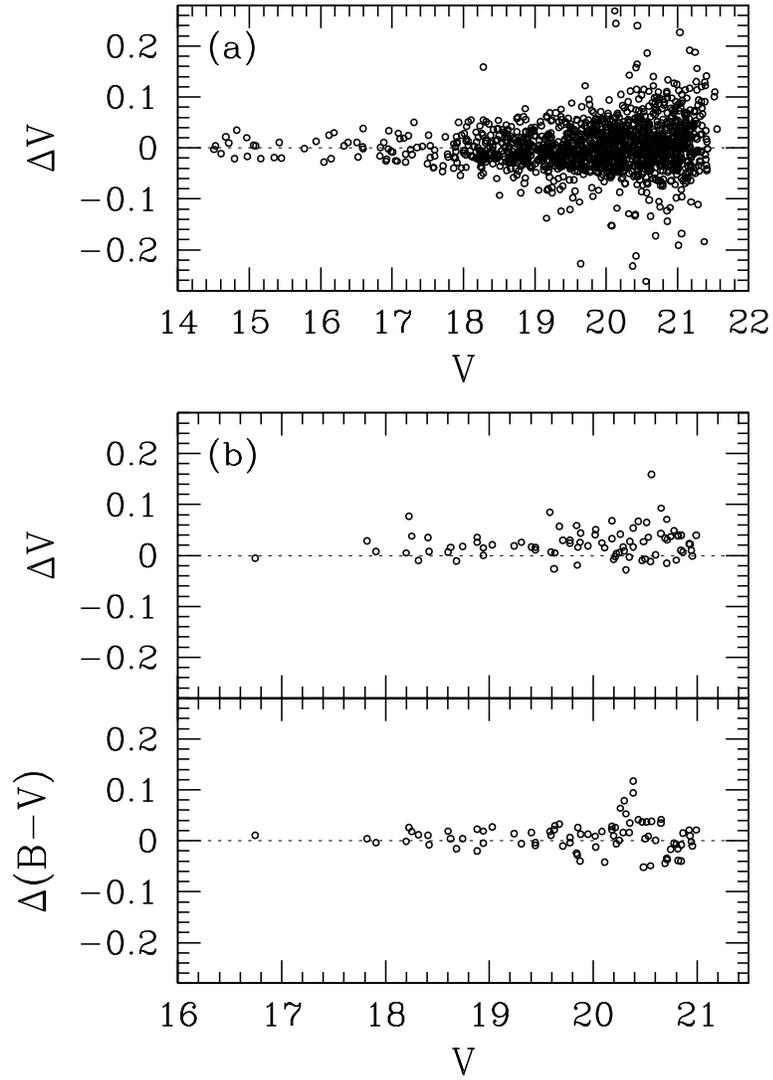}}
 \caption{Comparison of our photometry with (a) Johnson \& Bolte (1998) and (b) Richer \& Fahlman (1986)
for M13. The differences are in the sense of our photometry minus others.}
\end{figure}

\begin{figure}
\centerline{\epsfysize=8in%
\epsffile{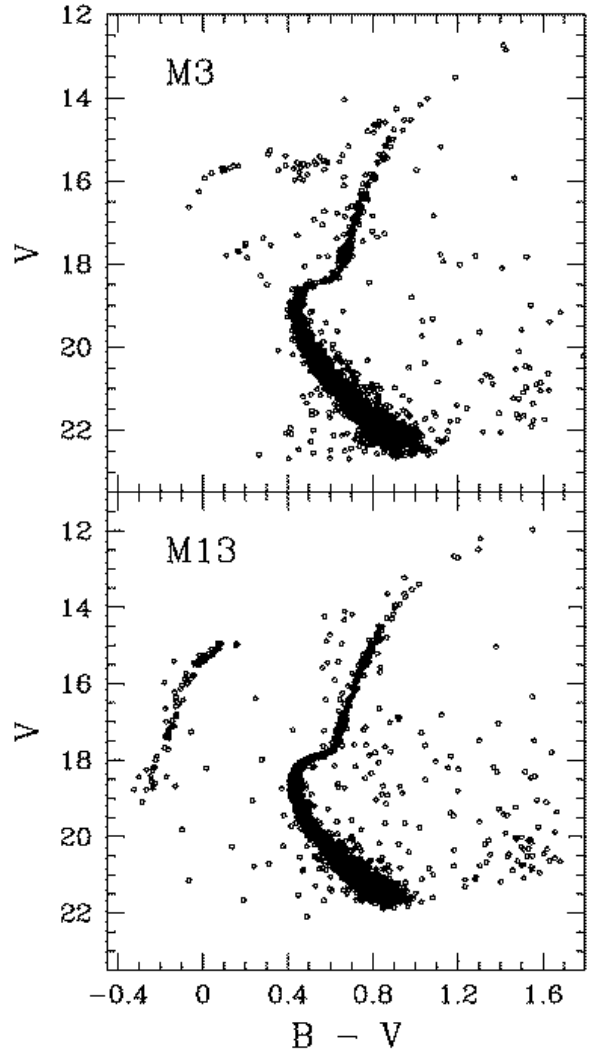}}
 \caption{($V, B-V$) CMDs for M3 (top) and M13 (bottom). See the text for the adopted selection criteria
that provide the best definitions of the CMD sequences.}
\end{figure}

\begin{figure}
\centerline{\epsfysize=7in%
\epsffile{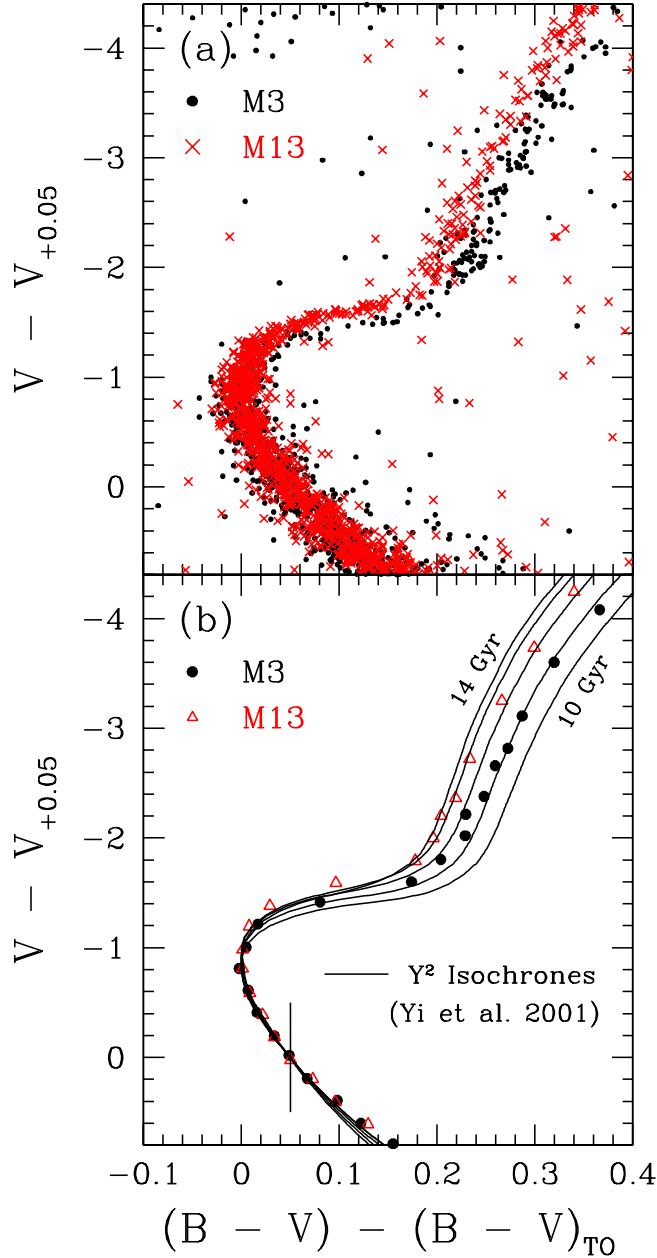}}
 \caption{(a) Comparison of our CM data of M3 (black filled circles) and M13
(red crosses), using the registration prescription specified by VBS.
(b) Same as (a) but for the fiducial sequences.
We also included Yonsei-Yale (Y$^{2}$) Isochrones (Yi et al. 2001)
having [Fe/H] = -1.66 and [$\alpha$/Fe] = 0.3 and ages ranging from 10 to 14 Gyr.
Note the large separation between the M3 and M13 RGBs, and slight offset in the
SGB region.}
\end{figure}

\begin{figure}
\centerline{\epsfysize=7in%
\epsffile{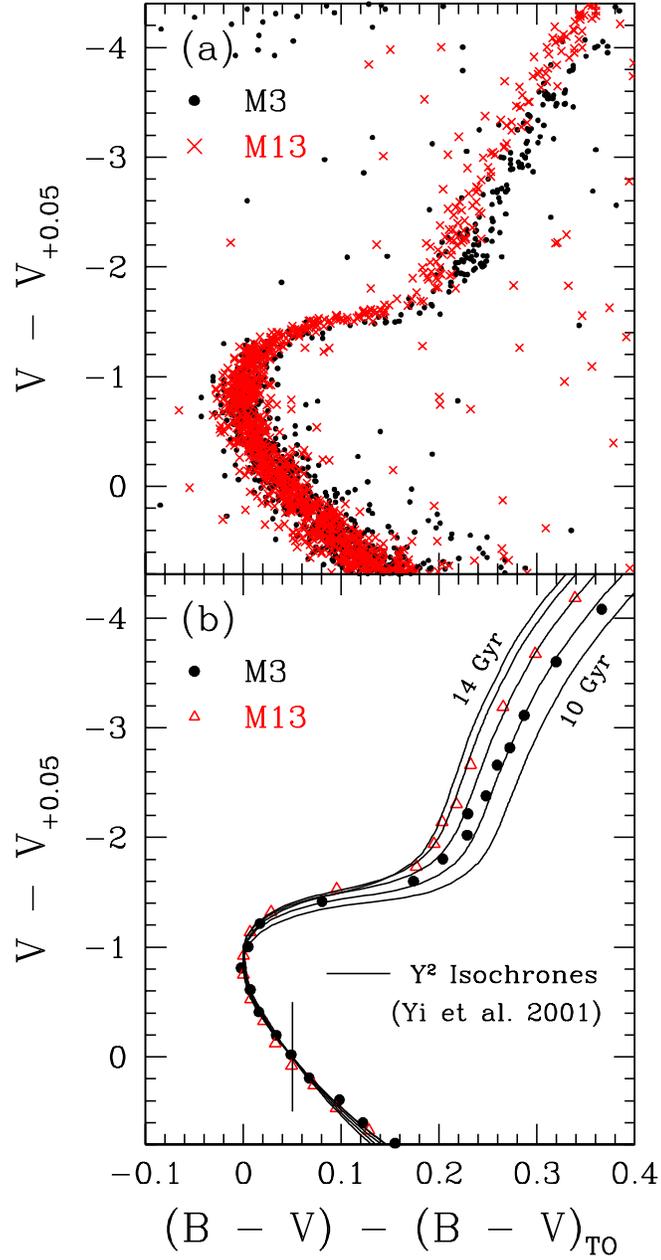}}
 \caption{Same as Fig. 7, but after the M13 sequence has been shifted in order to make the TOs
of both clusters coincide with each other (see text). Note that there is still a sufficiently
large separation between the M3 and M13 RGBs, implying a true age difference.}
\end{figure}

\begin{figure}
\centerline{\epsfysize=7in%
\epsffile{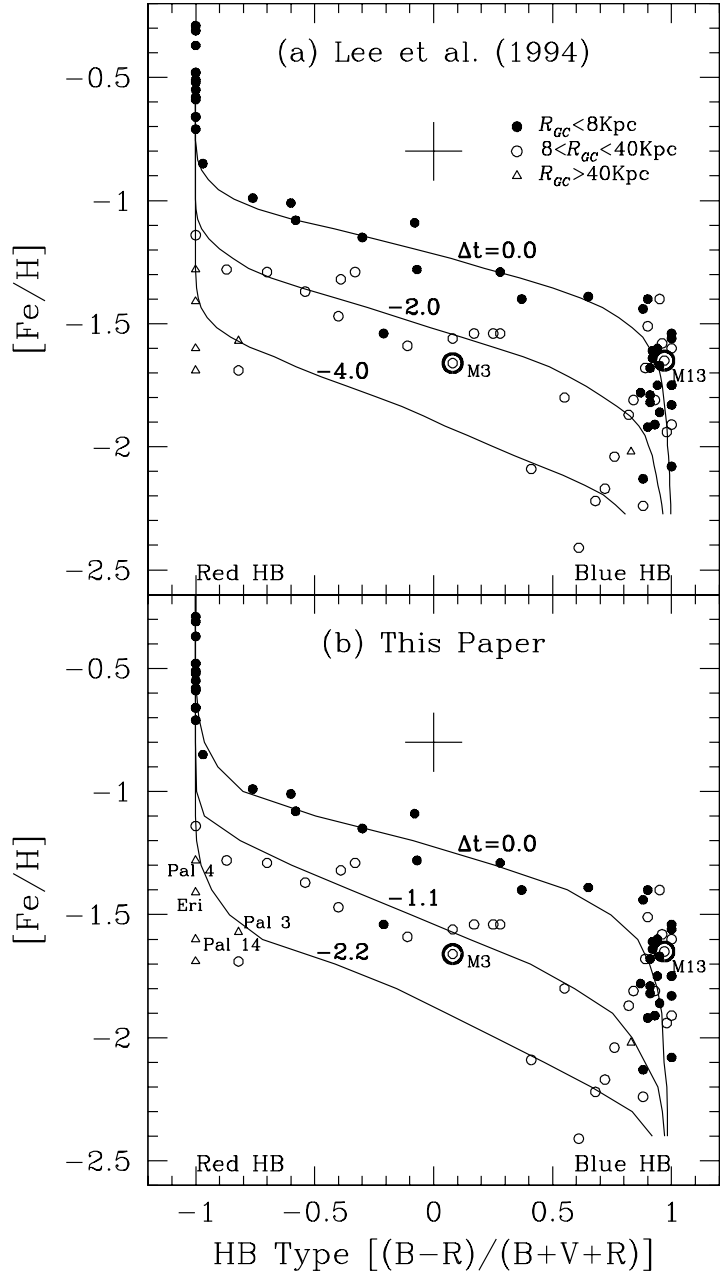}}
 \caption{HB morphology vs. [Fe/H] relations for Galactic globular clusters with
theoretical isochrones that were produced by synthetic HB models (cluster data from Table 1
of LDZ). Our new HB models with the effects of recent developments (b) are more
sensitive to age  compared to our earlier models (a). $\Delta t$ = 0 corresponds to
the mean age of the inner halo (R $<$ 8 kpc) clusters, and the relative ages are in Gyr.
Note that, with our new HB models, age differences of about 1 - 2 Gyr are
now enough to explain the observed differences in HB morphology between M3 and M13.}
\end{figure}

\begin{figure}
\centerline{\epsfysize=8in%
\epsffile{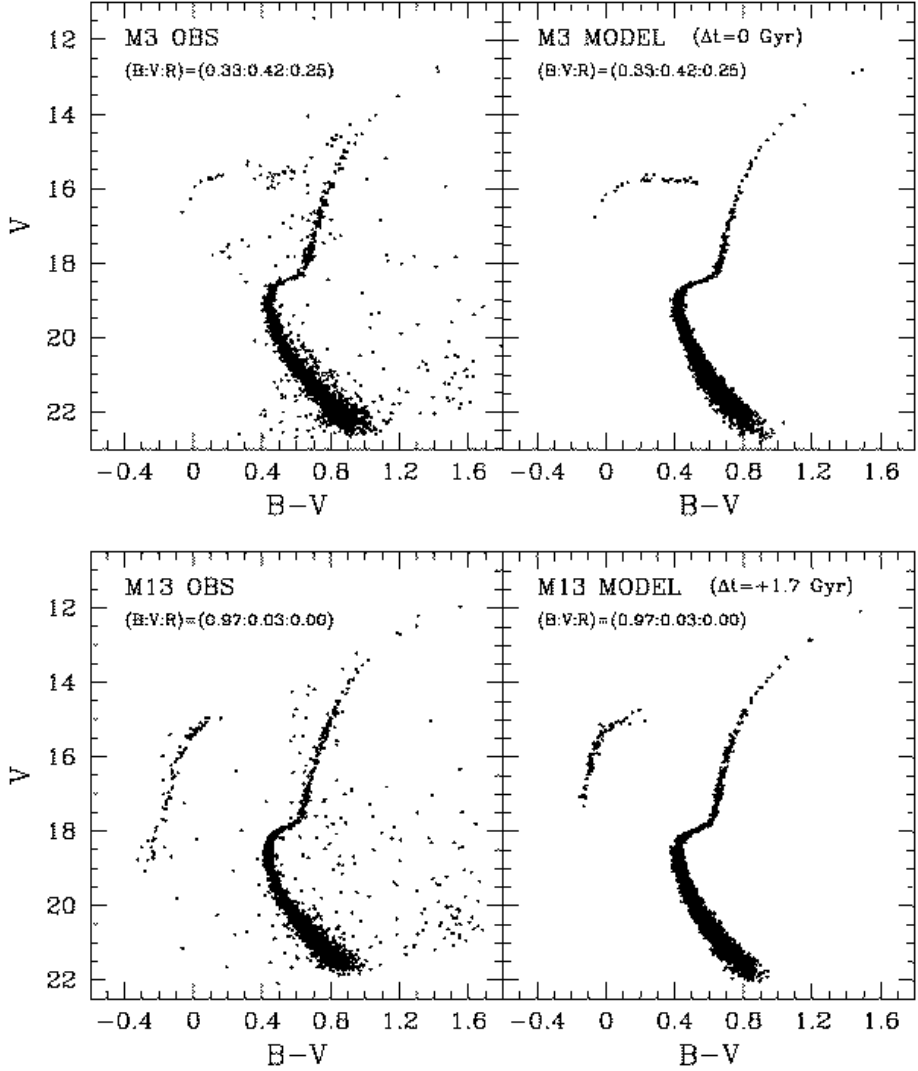}}
 \caption{Comparison of observational and synthetic CMDs for M3 and M13, assuming that
M13 is older by 1.7 Gyr. Crosses in the model CMDs are RR Lyrae variables.}
\end{figure}

\end{document}